\documentclass[]{aastex}
\usepackage[]{natbib}
\usepackage{graphics}
\usepackage{emulateapj5}
\usepackage{apjfonts}
 
\newcommand{\etal}{et al.}  
\newcommand{\per}{\ensuremath{^{-1}}}
\newcommand{\hbeta}{H\ensuremath{\beta}} 
\newcommand{\hst}{\emph{HST}}
\newcommand{\msun}{\ensuremath{M_{\odot}}}
\newcommand{\kms}{km s\ensuremath{^{-1}}} 
\newcommand{\mbh}{\ensuremath{M_\mathrm{\bullet}}}

\newcommand{\chisq}{\ensuremath{\chi^2}}
\newcommand{\sdssj}{{SDSS J1148+5251}}

\slugcomment{}
\shorttitle{IRON EMISSION IN SDSS J114816.64+525150.3} 
\shortauthors{BARTH ET AL.}

\begin{document} 

\title{Iron Emission in the $z=6.4$ Quasar SDSS J114816.64+525150.3}

\author{Aaron J. Barth\altaffilmark{1,2}, 
Paul Martini\altaffilmark{3},
Charles H. Nelson\altaffilmark{4},
and
Luis C. Ho\altaffilmark{3}
}

\altaffiltext{1}{Caltech Optical Observatories, 105-24 Caltech, Pasadena, CA
91125}

\altaffiltext{2}{Hubble Fellow}

\altaffiltext{3}{The Observatories of the Carnegie Institution of
  Washington, 813 Santa Barbara Street, Pasadena, CA 91101}

\altaffiltext{4}{Department of Physics and Astronomy, Drake
  University, 2507 University Avenue, Des Moines, IA 50311-4505}

\begin{abstract}

We present near-infrared $J$ and $K$-band spectra of the $z = 6.4$
quasar SDSS J114816.64+525150.3 obtained with the NIRSPEC spectrograph
at the Keck-II telescope, covering the rest-frame spectral regions
surrounding the \ion{C}{4} $\lambda1549$ and \ion{Mg}{2} $\lambda2800$
emission lines.  The iron emission blend at rest wavelength 2900--3000
\AA\ is clearly detected and its strength appears nearly
indistinguishable from that of typical quasars at lower redshifts.
The \ion{Fe}{2}/\ion{Mg}{2} ratio is also similar to values found for
lower-redshift quasars, demonstrating that there is no strong
evolution in Fe/$\alpha$ broad-line emission ratios even out to
$z=6.4$.  In the context of current models for iron enrichment from
Type Ia supernovae, this implies that the SN Ia progenitor stars
formed at $z \gtrsim 10$.  We apply the scaling relations of
Vestergaard and of McLure \& Jarvis to estimate the black hole mass
from the widths of the \ion{C}{4} and \ion{Mg}{2} emission lines and
the ultraviolet continuum luminosity.  The derived mass is in the
range $(2-6)\times10^9$ \msun, with an additional uncertainty of a
factor of 3 due to the intrinsic scatter in the scaling relations.
This result is in agreement with the previous mass estimate of
$3\times10^9$ \msun\ by Willott, McLure, \& Jarvis, and supports their
conclusion that the quasar is radiating close to its Eddington
luminosity.

\end{abstract}


\section{Introduction}

The broad emission lines of high-redshift quasars are luminous
beacons that can be used to study the metal enrichment history of
the densest regions in the early universe \citep[e.g.,][]{hf99}.
Observations of emission lines such as \ion{C}{4} and \ion{N}{5}
reveal supersolar abundances in quasar broad-line regions (BLRs) even
at $z>4$ \citep{die03a}, strengthening the link between the growth and
fueling of black holes and the buildup of stellar mass in the host
galaxies.  The abundance ratio of iron to $\alpha$ elements is a key
diagnostic that may be used as a ``clock'' to constrain the earliest
epoch of star formation \citep{hf93, ytk98}, since enrichment of
$\alpha$ elements occurs via core-collapse supernovae while the
dominant channel for iron enrichment is Type Ia supernovae, which have
longer-lived progenitor stars.

The flux ratio of the integrated iron emission in the 2200--3000 \AA\
wavelength range to \ion{Mg}{2} $\lambda\lambda2796, 2802$ is an
observationally accessible diagnostic that is dependent on the
underlying Fe/Mg abundance ratio, but also sensitive to the BLR
density and microturbulence \citep{ver03}.  Numerous studies have
demonstrated that there is no systematic decrease in the
\ion{Fe}{2}/\ion{Mg}{2} flux ratio of quasars with increasing
redshift, even out to $z\approx6$ \citep[e.g.,][]{the99, die02,
die03b, fck03}, while \citet{iwa02} found evidence for an
\emph{increase} in \ion{Fe}{2}/\ion{Mg}{2} with increasing redshift to
$z\approx5$.

SDSS J114816.64+525150.3 (abbreviated here as \sdssj) was discovered
by \citet{fan03} in the Sloan Digital Sky Survey; at $z=6.4$ it is the
highest-redshift quasar currently known.  \citet{wmj03} used $H$ and
$K$-band spectra of this object to obtain an estimate of the black
hole mass ($\sim3\times10^9$ \msun) from the \ion{Mg}{2} linewidth,
using the scaling relations derived by \citet{mj02}.  The \ion{Mg}{2}
$\lambda2800$ line was the only emission feature clearly present in
their spectrum; the limited S/N of their data did not permit a
definitive detection of the surrounding \ion{Fe}{2} emission blends.
The quasar environment was evidently a site of vigorous massive star
formation and metal enrichment prior to $z=6.4$, as demonstrated by
the roughly normal equivalent width of the \ion{Mg}{2} emission line
as well as the presence of submillimeter emission from $10^8$ \msun\
of dust in the quasar host galaxy \citep{ber03}.  Here, we present new
near-infrared spectra of \sdssj\ obtained at the Keck Observatory,
covering the rest-frame spectral regions surrounding the \ion{C}{4}
$\lambda1549$ and \ion{Mg}{2} $\lambda2800$ emission lines.  We show
that its \ion{Fe}{2}/\ion{Mg}{2} ratio is similar to that of typical
quasars at lower redshifts, and we update the black hole mass estimate
of Willott \etal\ using the higher-quality Keck data.  Except where
noted otherwise, we assume $H_0 = 70$ km s\per\ Mpc\per, $\Omega_m =
0.3$, and $\Omega_\Lambda = 0.7$; for this cosmology the universe at
$z=6.4$ was 840 Myr old.

\section{Observations and Reductions}

The observations were obtained using the NIRSPEC spectrograph
\citep{mcl00} at the Keck-II telescope on the nights of 2003 March 11
and 12 UT.  We used a $0\farcs76$-wide slit in the NIRSPEC-1, 2, and 6
settings, which cover the wavelength ranges 0.95--1.12, 1.08--1.29,
and 1.89--2.31 \micron, respectively, at $R \approx 1500$.  A standard
ABBA nodding sequence was used with exposure times of 300 s at each
nod position.  The airmass ranged from 1.2 to 1.5.  Seeing, as
measured from the spatial profiles of calibration star exposures, was
typically between 0\farcs7 and 0\farcs9.  The total exposure times in
each setting were 6000, 9600, and 8400 s, respectively.  Due to
hardware problems, the nodding and guiding performance of NIRSPEC was
relatively poor during this run and slit losses were severe, possibly
$\sim50\%$ during some individual exposures.

The spectra were extracted from bias-subtracted, flattened
2-dimensional frames using an optimal weighting algorithm, and using
the sky-subtraction code of \citet{kel03}.  Wavelength calibration was
performed with the OH airglow emission lines in each exposure.  The
data were flux-calibrated and corrected for telluric absorption using
spectra of the A0V star HD 99966, observed immediately before or after
the quasar in each setting, following the methods described by
\citet{vcr03}.  Finally, the spectra were scaled to the $J$ and
$K^{\prime}$ magnitudes given for \sdssj\ by \citet{fan03}, to place
them on an absolute flux scale.

\section{Results and Discussion}

\begin{figure}
\plotone{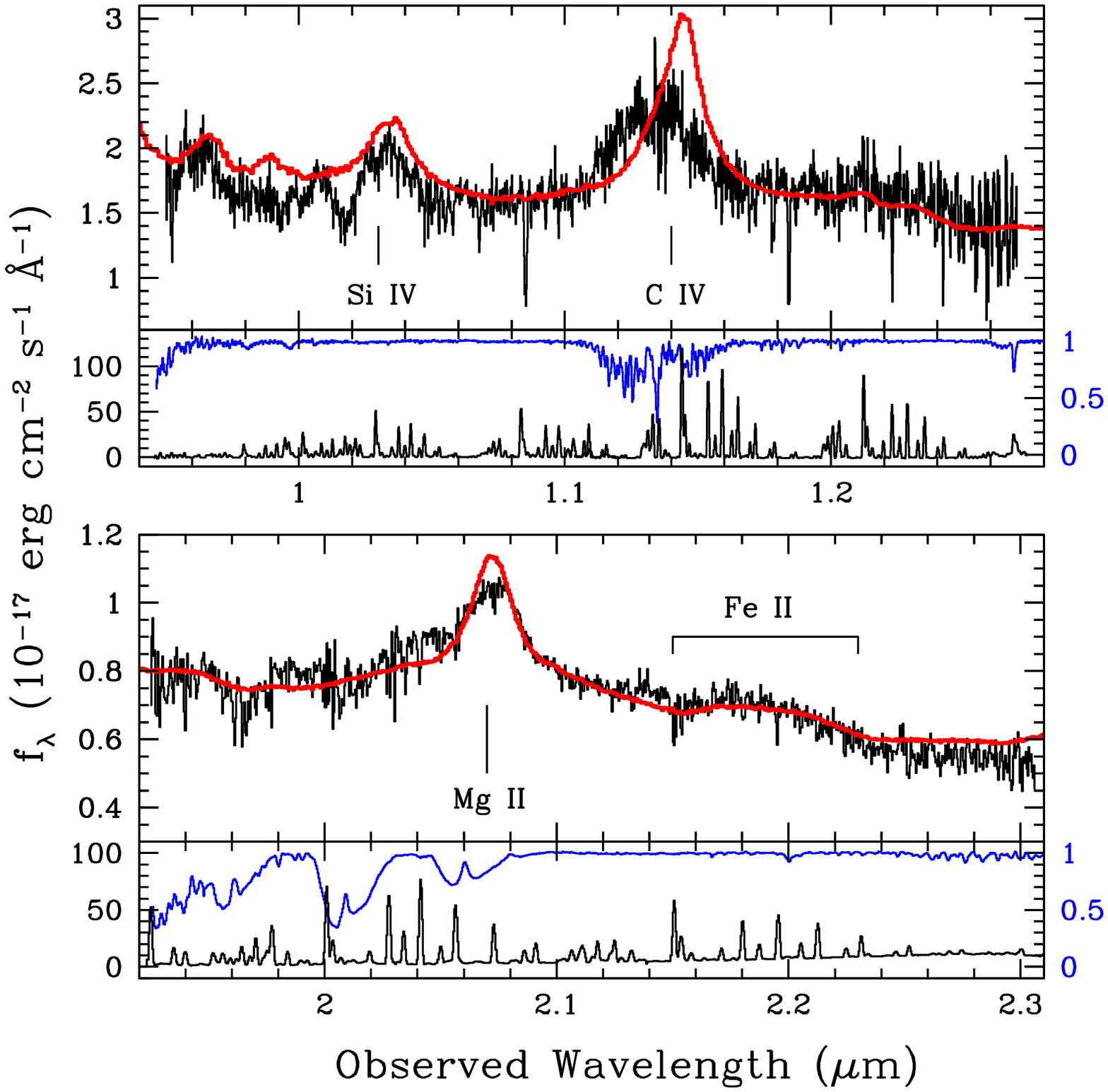}
\caption{Near-infrared spectra of \sdssj.  The upper panels show the
  combined NIRSPEC-1 and 2 settings, and the lower panels show the
  NIRSPEC-6 setting.  The spectrum overplotted on that of \sdssj\ is
  the SDSS composite quasar of \citet{vdb01}, transformed to $z=6.40$.
  The same scaling factor has been applied to the SDSS composite
  quasar in both spectral regions.  The panels below the quasar
  spectra show the atmospheric transmission and the night sky emission
  spectra.  The Si IV, C IV, and Mg II emission lines are labelled.
  There is iron emission throughout this spectral region; the
  prominent iron blend at 2900--3000 \AA\ rest wavelength is
  labelled.}
\label{obsframe}
\end{figure}

The spectra are displayed in Figure \ref{obsframe}, overplotted with
the SDSS composite quasar spectrum of \citet{vdb01}.  The best match
between the \ion{Mg}{2} profile of SDSS J1148+5251 and the SDSS
composite is found for a redshift of $6.40 \pm 0.01$, in agreement
with the previous measurement of $z = 6.41\pm0.01$ by \citet{wmj03}.
However, the \ion{C}{4} emission line appears strongly blueshifted
with respect to the \ion{Mg}{2} redshift and is much broader than the
\ion{C}{4} line of the composite quasar.  The \ion{C}{4} line falls
within a telluric absorption band at 1.14 \micron, making accurate
flux calibration difficult, but the extended blue wing of \ion{C}{4}
does appear to be a real feature in the spectrum.  In quasars,
\ion{Mg}{2} is considered a good indicator of the systemic redshift
while relative \ion{C}{4} blueshifts of 500--2000 \kms\ or more are
common \citep{mar96, ric02}.  The centroid of the \ion{C}{4} line
measured by a Gaussian fit corresponds to a blueshift of $2900 \pm
120$ \kms\ relative to the \ion{Mg}{2} line.  \citet{ric02} find that
the \ion{Si}{4} $\lambda1400$ emission line of quasars is typically
not strongly blueshifted, unlike \ion{C}{4}; Figure \ref{obsframe}
shows that \sdssj\ appears to follow this trend.

The data do not reveal any broad absorption lines, but a few narrow
absorption lines are present.  The strongest are at 1.0853 \micron\
(equivalent width $5.6 \pm 0.8$ \AA) and at 1.1842 \micron\
(equivalent width $4.4 \pm 0.4$ \AA).  Neither is coincident with an
atmospheric absorption or emission feature, and they appear to be
genuine features in the quasar spectrum.  The line at 1.1842 \micron\
is unresolved, while the 1.0853 \micron\ feature is marginally broader
than the night sky emission lines but not broad enough to be
consistent with a blended \ion{C}{4} or \ion{Mg}{2} doublet.  The
identification of these lines is uncertain, and neither of them
matches any wavelengths that would be expected for metal absorption
lines from the $z=4.943$ \ion{C}{4} absorber found by \citet{whi03}.

\subsection{\ion{Fe}{2} Emission}

In general, measurement of the iron emission strength is best
accomplished when the data have wide spectral coverage, so that
spectral regions without either iron emission or Balmer continuum
emission can be included in the fit \citep[e.g.,][]{die02}.  Our
spectrum does not extend to sufficiently long wavelengths to do this,
and we fit only the rest-frame spectral region 2610--3100 \AA.  An
\ion{Fe}{2} emission blend at 2900--3000 \AA\ is clearly present
in the Keck spectrum.  From the comparison between the NIRSPEC data
and the SDSS composite spectrum in Figure \ref{obsframe}, it is
already apparent that the \ion{Fe}{2}/\ion{Mg}{2} ratio in SDSS
1148+5251 is roughly similar to that in the composite quasar.

The spectrum was modeled as the sum of three components: a power-law
continuum, an empirical iron emission template, and a double-Gaussian
model for the \ion{Mg}{2} profile.  For the \ion{Mg}{2} model, all of
the Gaussian parameters were allowed to float freely.  We do not
ascribe any particular physical interpretation to the two components
in terms of the structure of the emission-line regions; this is simply
the minimal empirical description that allows an adequate fit to the
emission line.  \citet{die02} show that the \ion{Mg}{2} profiles of
quasars can generally be fit well by a Gaussian near the systemic
velocity plus another blueshifted (and usually broader) Gaussian
component.  We note that the model fit may be affected by an imperfect
correction for telluric absorption in the blue wing of \ion{Mg}{2}.
Given our limited spectral coverage we chose not to include the Balmer
continuum emission as a separate component, since its strength could
not be determined independently of the dominant power-law continuum.

\begin{figure}
\plotone{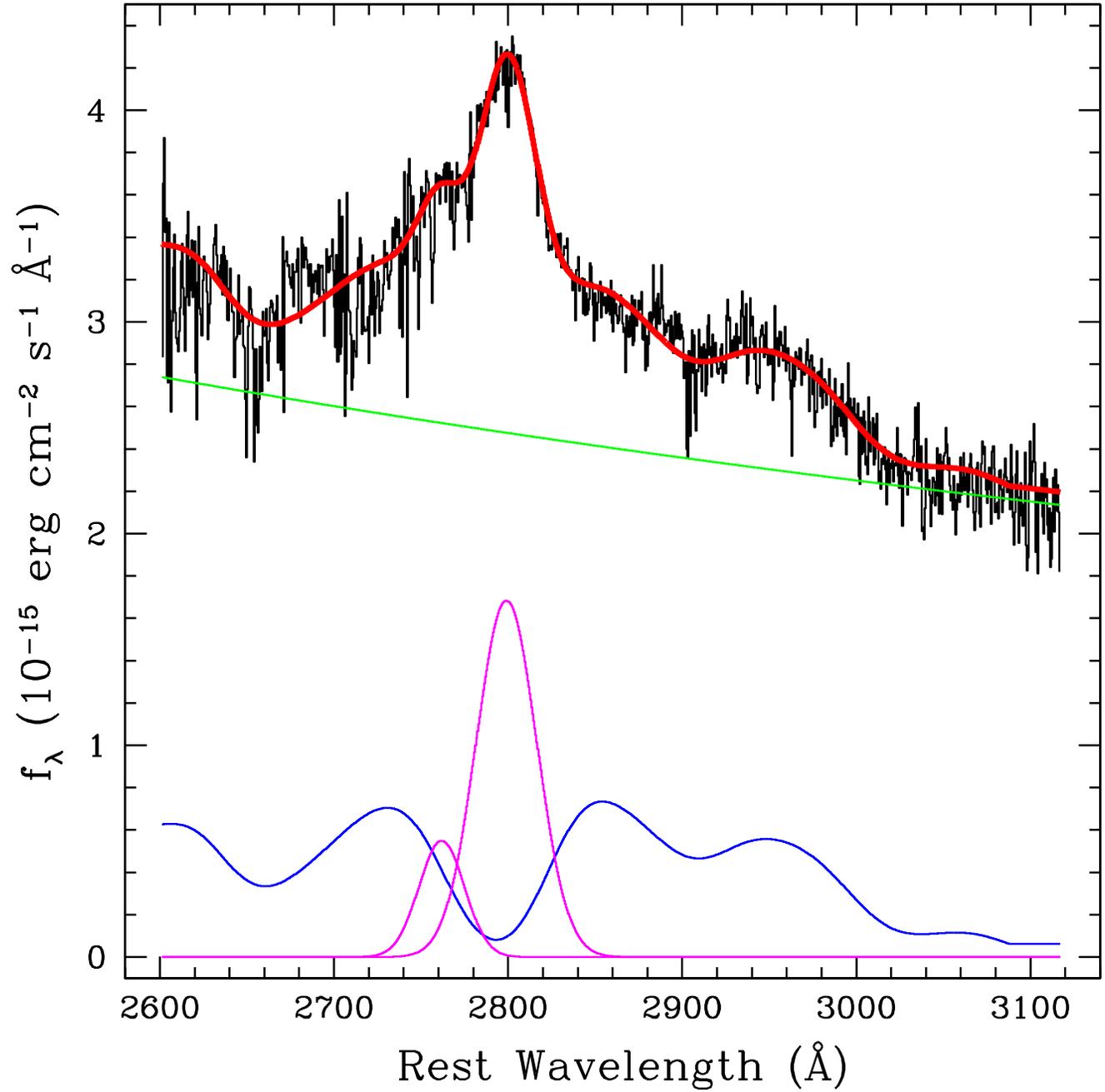}
\caption{Model fit to the Mg II spectral region.  The double-Gaussian
    model for the Mg II profile, the broadened Fe emission template,
    and the power-law continuum are displayed individually, and the
    full model is overplotted on the quasar spectrum.  The observed
    spectrum has been shifted to the rest frame and the flux density
    has been scaled by $(1+z)^3$. }
\label{feplot}
\end{figure}

The iron blends in this wavelength region are traditionally modeled
using an empirical ``template'' spectrum derived from an AGN with
intrinsically narrow emission lines.  We used an iron template derived
from \hst\ ultraviolet observations of I~Zw~1 as described by
\citet{vw01}\footnote{The \citet{vw01} iron template is not
distributed or shared freely, so we re-created their template spectrum
using information given in their published article.}; the iron
emission is dominated by \ion{Fe}{2} blends with a contribution from
\ion{Fe}{3}.  The iron template was broadened by convolution with a
Gaussian having the same width as the stronger \ion{Mg}{2} component.
To optimize the fit we computed a \chisq\ parameter determined using
the $1\sigma$ uncertainties on each pixel in the flux-calibrated
spectra.  Figure \ref{feplot} displays the best-fitting model.
The power-law index of the model continuum is $\alpha = -1.4 \pm 0.1$
for $f_\lambda \propto \lambda^\alpha$, similar to the value of
$\alpha = -1.56$ for the SDSS composite quasar \citep{vdb01}.

Following \citet{die02}, the \ion{Fe}{2}/\ion{Mg}{2} emission-line
ratio was determined by integrating the iron template flux over the
wavelength range 2200--3090 \AA.  Although this requires extrapolation
of the scaled iron template beyond the wavelength range over which our
fit was performed, it allows the most straightforward comparison with
previous measurements.  We find \ion{Fe}{2}/\ion{Mg}{2} = $4.7 \pm
0.4$.  This result is similar to values that have been found for
quasars over all observed redshifts.  For example, \citet{die02}
obtained \ion{Fe}{2}/\ion{Mg}{2} = 3.3--4.2 for six quasars at
$z\approx 3.4$ and \ion{Fe}{2}/\ion{Mg}{2} = $3.8 \pm 0.4$ for a
low-$z$ composite quasar, while \citet{the99} measured
\ion{Fe}{2}/\ion{Mg}{2} $\approx 6-7$ (over a slightly wider
wavelength range of 2000--3000 \AA\ for \ion{Fe}{2}) for composite
quasar spectra at $z=3$ and $z=4$, and $4.3\pm0.1$ for a low-$z$
composite quasar.  Our result is also within the same range as values
measured by \citet{iwa02}, \citet{die03b}, and \citet{fck03} for
quasars at $z = 4.4-6.3$.

Thus, the lack of any strong evolution in \ion{Fe}{2}/\ion{Mg}{2}
noted by Thompson \etal\ out to $z= 3-4$ apparently continues out to
$z=6.4$; even at this redshift there is still no indication of a
substantial decrease in this ratio.  If the \ion{Fe}{2}/\ion{Mg}{2}
flux ratio can be interpreted as even a rough indicator of the
underlying abundance ratio, then the age of the universe at $z=6.4$
provides some constraints on metal enrichment models.  The minimum
lifetime for SN Ia progenitors has often been considered to be
$\sim1$--1.5 Gyr; this would imply that the Fe/$\alpha$ abundance
ratio should undergo a strong rise starting at 1--1.5 Gyr following
the first burst of star formation in which SN Ia progenitors are
created \citep{hf93, ytk98}.  

Observations of \ion{Fe}{2}/\ion{Mg}{2} ratios in quasars at
$z\approx4$ could still be consistent with such models \citep{die02},
but our observations and those of \citet{fck03} rule out models with
iron enrichment timescales of $\gtrsim800$ Myr.  However, the
timescale for the maximum SN Ia rate (and hence iron enrichment)
depends sensitively on the star formation rate and initial mass
function, and may be much shorter than the canonical 1~Gyr for an
intense starburst in the core of a rapidly forming elliptical galaxy
\citep{ft98, mr01}.  The models of \citet{ft98} give a typical
timescale of 0.3 Gyr to reach solar abundance of iron in the
interstellar medium of a proto-elliptical galaxy; similarly,
\citet{mr01} find that the peak SN Ia rate in an elliptical galaxy
occurs $\sim0.3$ Gyr after the major burst of star formation.  If the
iron abundance were built up in just 0.3 Gyr, then the SN Ia
progenitors would have formed at $z \gtrsim 10$.  There may be only a
brief temporal window during which a quasar would exhibit a low
Fe/$\alpha$ abundance ratio, and this epoch could be at a redshift as
high as $\sim10$ if star formation began at $z \approx 20$
\citep[e.g.,][]{kog03}.  Alternatively, the iron may have been
produced in pair-instability supernovae from Population III stars with
initial masses of 140--260 \msun; a single such explosion could
produce up to 40 \msun\ of iron \citep{hw02}.  This would result in
iron enrichment of the pregalactic gas within a few Myr after the
first burst of star formation.

\subsection{The Mass of the Black Hole}

The only practical methods to estimate the black hole masses in
distant quasars are based on scaling relations derived from
low-redshift samples.  Reverberation mapping of Seyfert galaxies and
quasars at $z < 0.4$ has demonstrated a correlation between continuum
luminosity and BLR radius \citep{wpm99,kas00}.  By combining the
continuum luminosity with measurements of broad-line widths, the black
hole mass can be estimated under the assumption of virial motion of
the BLR clouds.  The original scaling relations were derived from
FWHM(\hbeta) and the continuum luminosity at 5100 \AA, but similar
relations have been determined using ultraviolet continuum luminosity
combined with the linewidth of either \ion{C}{4} \citep{ves02} or
\ion{Mg}{2} \citep{mj02}.  These relations have the form $\mbh \propto
[\lambda L_\lambda$(UV)$]^\beta \times$ FWHM(line)$^2$, where $\beta$
= 0.47 \citep{mj02} or 0.7 \citep{ves02}.  The dominant uncertainty in
the \mbh\ estimates comes from the intrinsic scatter in the
\mbh--linewidth correlations, which is a factor of $\sim2.5-3$, rather
than from measurement uncertainties on the linewidths.

\citet{wmj03} found FWHM(\ion{Mg}{2}) = $6000^{+1100}_{-600}$ \kms\
based on a double-Gaussian fit with the two components fixed at rest
wavelengths of 2796 and 2802 \AA, giving $\mbh \approx 3\times10^9$
\msun.  Our spectrum shows that the profile is rather asymmetric, and
the full \ion{Mg}{2} profile has FWHM(\ion{Mg}{2}) = $5500 \pm 200$
\kms.  Applying the \citet{mj02} relation with $\lambda
L_{\mathrm{\lambda}}$(3000 \AA) $= 5.7 \times 10^{39}$ W, we obtain
$\mbh \approx 2\times10^9$ \msun.  This result is close to the
previous estimate by Willott \etal, since we find similar values for
FWHM(\ion{Mg}{2}) and $L_\lambda$(3000 \AA).

The peculiar shape of the \ion{C}{4} line might result from
non-Keplerian kinematics, calling into question the application of the
\citet{ves02} method to derive \mbh\ for this particular object.
Keeping this caveat in mind, we will derive an \mbh\ estimate from the
\ion{C}{4} line for purposes of comparison with the \ion{Mg}{2}
method.  To measure FWHM(\ion{C}{4}), the spectrum was first prepared
by subtracting off the Fe template, with its overall scaling
determined by the fit to the $K$-band spectrum.  The Fe emission only
makes a small ($\sim10\%$) contribution to the flux level in the
continuum surrounding \ion{C}{4}.  Then, the region from 1450 to 1630
\AA\ rest wavelength was fit with a single Gaussian plus continuum.
The best-fit Gaussian has FWHM(\ion{C}{4}) = $9000 \pm 300$ \kms.
Vestergaard's \mbh\ relation was derived assuming a cosmology with
$H_0 = 75$ km s\per\ Mpc\per, $q_0 = 0.5$, and $\Lambda = 0$.  Taking
$\lambda L_{\lambda}$(1350 \AA) $= 2.5\times10^{46}$ erg s\per\ for
these cosmological parameters, we obtain $\mbh \approx 6\times10^9$
\msun\ for \sdssj.  Given that these scaling methods have an estimated
uncertainty of a factor of 2.5--3, the two estimates of \mbh\ from
\ion{Mg}{2} and \ion{C}{4} are in reasonable agreement.  It would
still be worthwhile to obtain a \ion{C}{4} spectrum with higher S/N
ratio to check the reality of the large blueshift and linewidth.

To estimate the bolometric luminosity, we apply a bolometric
correction of $L_\mathrm{bol}/[\nu L_\nu(2500$~\AA$)] = 6.3$, which is
typical of low-redshift quasars \citep{elv94}.  This gives
$L_\mathrm{bol} \approx 4\times10^{47}$ erg s\per\ and implies (as
previously shown by Willott \etal) an Eddington ratio of
$L_\mathrm{bol}/L_\mathrm{Edd} \approx 1$ for $\mbh = 3\times10^9$
\msun.

A few additional caveats should be noted: the apparent luminosity of
the quasar can be affected by dust extinction, continuum beaming, or
gravitational lensing. Significant extinction is unlikely because the
spectral shape of the quasar does not appear strongly reddened
relative to the SDSS composite quasar, while \citet{wmj03} argue
against continuum beaming because the \ion{Mg}{2} equivalent width is
typical of unbeamed quasars.  Lensing is expected to affect only a
small fraction of high-$z$ SDSS quasars \citep{wl02, pin03}, and
infrared images of \sdssj\ do not give any indication of lensing
\citep{fan03}.  

It is remarkable to find such a massive black hole at such a large
look-back time.  Nevertheless, models for the hierarchical
growth of supermassive black holes by merging and gas accretion can
accommodate masses of $>10^9$ \msun\ in quasars at $z \gtrsim 6$ if
the seed black holes form at $z>10$ \citep[e.g.,][]{hl01, vhm03}.  The
initial seed black holes may be objects with $\mbh \approx 100-200$
\msun\ formed by collapse of supermassive Population III stars
\citep{fwh01, mad01, sch02}.  We note that even a single seed black
hole having 180 \msun\ at $z=20$ could (just barely) grow to
$3\times10^9$ \msun\ by $z=6.4$ if it were constantly accreting at the
Eddington luminosity with a radiative efficiency of $\epsilon = 0.1$,
so the present data do not yet \emph{require} multiple black hole
mergers to build up the observed black hole mass.

\section{Conclusions}

The Keck spectra reveal an \ion{Fe}{2}/\ion{Mg}{2} ratio in \sdssj\
that is within the normal range for quasars at lower redshifts,
indicating that substantial iron enrichment must have occurred in the
densest regions of the universe at redshifts greater than 6.4.  Recent
photoionization calculations by \citet{ver03} have shown that
\ion{Fe}{2}/\ion{Mg}{2} is not a straightforward diagnostic of
Fe/$\alpha$ abundance ratios, but useful constraints on the
Fe/Mg abundance ratio can be derived from observations of the
ultraviolet \ion{Fe}{2}/\ion{Mg}{2} ratio combined with the ratio of
ultraviolet to optical \ion{Fe}{2} strength.  The rest-frame optical
\ion{Fe}{2} emission in very high-redshift quasars cannot be measured
from the ground, but these measurements can be pursued in the future
with the \emph{James Webb Space Telescope}.  Searches for quasars at
still higher redshifts may provide a glimpse of the earliest epoch of
iron production and give stringent constraints on the formation
redshift of the seed black holes that were the progenitors of luminous
quasars.

\acknowledgments 

We thank D. Kelson for the use of his sky-subtraction code and
D. Stern for helpful comments.  Research by A.J.B. is supported by
NASA through Hubble Fellowship grant \#HST-HF-01134.01-A awarded by
STScI.  P.M. was supported by a Carnegie Starr Fellowship.  Data
presented herein were obtained at the W.M. Keck Observatory, which is
operated as a scientific partnership among Caltech, the University of
California, and NASA. The Observatory was made possible by the
generous financial support of the W.M. Keck Foundation.  The authors
wish to recognize and acknowledge the very significant cultural role
and reverence that the summit of Mauna Kea has always had within the
indigenous Hawaiian community.

\end{document}